\documentclass[twocolumn,pre]{revtex4}

\usepackage{dcolumn}
\usepackage{amsmath}
\usepackage{amssymb}
\usepackage{txfonts}

\usepackage{graphicx}

\begin{document}

\renewcommand{\d}{d}
\newcommand{\Ord}{\mathrm{O}}
\newcommand{\e}{\mathrm{e}}
\newcommand{\ii}{\mathrm{i}}
\newcommand{\half}{\mbox{$\frac12$}}
\newcommand{\set}[1]{\lbrace#1\rbrace}
\newcommand{\av}[1]{\left\langle#1\right\rangle}
\newcommand{\eref}[1]{(\ref{#1})}
\newcommand{\etal}{{\it{}et~al.}}
\newcommand{\defn}{\textit}
\newcommand{\mat}{\mathbf}
\renewcommand{\vec}{\mathbf}

\newlength{\figurewidth}
\setlength{\figurewidth}{0.95\columnwidth}
\setlength{\parskip}{0pt}
\setlength{\tabcolsep}{6pt}
\setlength{\arraycolsep}{2pt}

\title{Embedded-Graph Theory}
\author{Atsushi Yokoyama}
\affiliation{}

\begin{abstract}
In this paper, we propose a new type of graph, denoted as "embedded-graph", and its theory, which employs a distributed representation to describe the relations on the graph edges. Embedded-graphs can express linguistic and complicated relations, which cannot be expressed by the existing edge-graphs or weighted-graphs. We introduce the mathematical definition of embedded-graph, translation, edge distance, and graph similarity. We can transform an embedded-graph into a weighted-graph and a weighted-graph into an edge-graph by the translation method and by threshold calculation, respectively. The edge distance of an embedded-graph is a distance based on the components of a target vector, and it is calculated through cosine similarity with the target vector. The graph similarity is obtained considering the relations with linguistic complexity. In addition, we provide some examples and data structures for embedded-graphs in this paper.
\end{abstract}
\maketitle

\section{Introduction}
Various systems can be represented using graph theory, for instance, Internet networks, social networks, electrical circuits, and biochemical pathways \cite{Strogatz01,AB02,DM02}. With the introduction of graph convolution, large-scale graphs can be analyzed using new deep learning technologies, that can be applied to many types of networks \cite{KW,DX}. 

\begin{figure}
\begin{center}
\resizebox{\figurewidth}{!}{\includegraphics{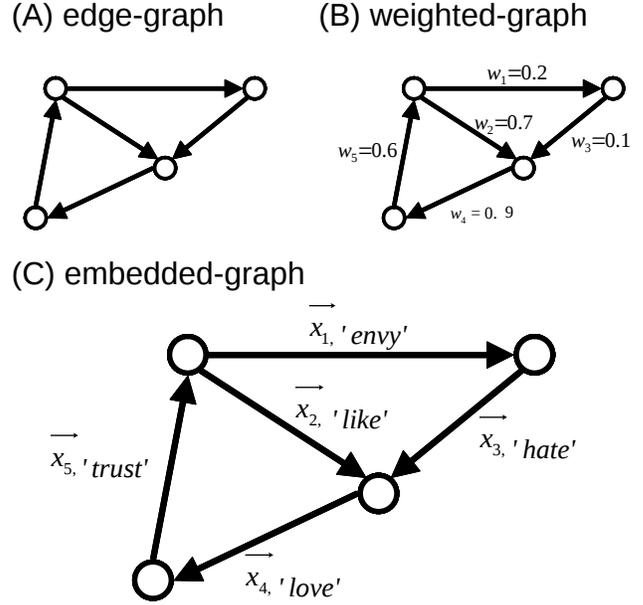}}
\end{center}
\caption{Examples of (A) edge-graph, (B) weighted-graph, and (C) embedded-graph, which is proposed in this paper. }
\label{F1}
\end{figure}

To use graph analysis certain limitations need to be resolved, such as poor representative capability. The most common type of graph, which is also the one stemming from the original definition, is the edge-graph, where pairs of vertices are linked by edges. An edge-graph is a simple structure, and has been studied extensively in the past \cite{BA}; however, it cannot represent complicated relations such as "like" or "love", because it can express only the existence of relations between pairs of vertices . An example of edge-graph is shown in Fig. \ref{F1}(A).

In a weighted-graph, edges have an associated value that expresses the strength of relation \cite{GR}. However, it is difficult to describe the difference between "like", "trust", "approve", or "hate", "envy", "doubt", because a weighted-graph can represent only the strength of relation and not the broad variety of expressions that natural language or words have or the subtle nuances between words. An example of weighted-graph is shown in Fig. \ref{F1}(B).

Recently, since the introduction of word2vec, many types of vector representations have been investigated, such as image2vec and doc2vec. Word2vec is a model to produce continuous vector representations of words that carry semantic meanings, and it is used to measure similarity of words, or operate word vectors mathematically, an example of which, [king] - [man] + [woman] ~= [queen], is widely known \cite{GE,TC,MW,MW,ME,ME02}. Word2vec can represent many of words using only tens to hundreds of vectors.

In this paper, we propose a new graph representation, called embedded-graph, which can use embedding representations such as word2vec to describe complicate relations between nodes, such as "like", "trust", or "approve", which cannot be expressed by the existing types of graphs. An example of embedded-graph is shown in Fig. \ref{F1}(C).

Embedded-graphs can be translated into weighted-graphs by certain mathematical operations explained in the following section, and furthermore, weighted-graphs can be translated into edge-graphs by a simple threshold calculation. Specifically, there is a smooth downward relation between embedded-graphs, weighted-graphs, and edge-graphs.

We explain the mathematical definition, translation theory, and other calculations related to embedded-graphs in Section \ref{sec:Theory}. We present examples of applications and experiments in Section \ref{sec:Application}, and the data structures used to represent embedded-graphs in Section \ref{sec:Data}. Finally, we discuss embedded-graphs including their drawbacks in Section  \ref{sec:Discussion}.

\section{Theory}
\label{sec:Theory}
\subsection{Definition}

An edge-graph is represented as $G=(V,E)$ where $V$ is the set of vertices, $E \subset  V \times V$ is the set of edges, and both are represented as vectors,

\begin{eqnarray}
  V =  \left\{ v_1, v_2, ..., v_{N_v} \right\} \nonumber \\
  E = \left\{ e_1, e_2, ..., e_{N_e} \right\},  \nonumber
\end{eqnarray}

\noindent where $N_v$ is the number of vertices and $N_e$ is the number of edges.

A weighted-graph is represented as $G=(V,E,W)$ where $W : E  \to \mathbb{R}$ assigns a weight to each edge $e \in E$. Map $W$ can be written as 

\begin{eqnarray}
  W(e) =  w, \nonumber 
\end{eqnarray}

\noindent  where $w$ is a scholar.

An embedded-graph is expressed as $G=(V,E,X)$ where $X : E  \to \mathbb{R}^d$ assigns a vector to each edge, which represents a word, a doc, an object, and so on. Here $d$ stands for the vector dimension. Map $X$ can be written as 

\begin{eqnarray}
  X(e) =  \left\{ x_1, x_2, ..., x_{d} \right\}. \nonumber 
\end{eqnarray}

When word2vec is employed as the embedded vector, $X$ becomes

\begin{eqnarray}
  X(e) =  \left\{ x_1, x_2, ..., x_{d} \right\} = \overrightarrow{x}_{word} \nonumber 
\end{eqnarray}

\noindent where $\overrightarrow{x}$ denotes the word vector. For instance, the word 'king' is represented by $\overrightarrow{x}_{'king'}$.


\subsection{Translation}
The three types of graphs, edge-graphs, weighted-graphs and embedded-graphs, can be connected by certain mathematical operations. The translation function from the embedded vector to the weight scalar is the function

\begin{eqnarray}
  w = F(X(e)). \nonumber 
\end{eqnarray}

\noindent  For example, the following function, denoted by cosine similarity, can be defined as the translation function

\begin{eqnarray}
  F(X(e); X^*) \coloneqq cos(X(e), X^*) \label{eq:cos}
\end{eqnarray}

\noindent  which measures the similarity between vectors $X(e)$ and $X^*$, where $X^*$ is a selected word vector. For instance, when an embedded-graph needs to be investigated from the point of view of 'reliability', $X^*$ becomes $X^*=\overrightarrow{x}_{'reliability'}$, and all relationships between vertices are calculated by Eq. \ref{eq:cos} to obtain the weight scalar number for each edge. In other words, all similarities between the word vector on each edge and the target word 'reliability' are obtained and a value of 'reliability' for each edge is calculated. In this manner, an embedded-graph is translated into a weighted-graph. 

\begin{figure}
\begin{center}
\resizebox{\figurewidth}{!}{\includegraphics{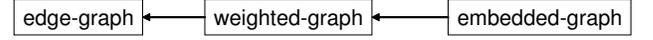}}
\end{center}
\caption{Translation between an edge-graph, a weighted-graph, and an embedded-graph. An embedded-graph can be translated into a weighted-graph by taking the cosine distance or the inner product with a specific vector. A weighted-graph can be translated into an edge-graph by threshold calculation.}
\label{F2}
\end{figure}

It is known that a weighted-graph can be translated into an edge-graph by a simple threshold calculation and therefore our proposed embedded-graph, the weighted-graph, and the edge-graph can only be translated downwards, as shown in Fig \ref{F2}. 

The function defined in Eq.  \ref{eq:cos} is not the only way to obtain a scalar from vectors: other functions can be also used, such as the one defined in Eq. \ref{eq:inner}. 

\begin{eqnarray}
  F(X(e); X^*) \coloneqq X(e) \cdot X^* \label{eq:inner}
\end{eqnarray}

\subsection{Edge distance}

The distance between adjacent vertices can also be defined using cosine similarity as

\begin{eqnarray}
  d(e, X^*) \coloneqq  1 - cos(X(e), X^*). \label{eq:d}
\end{eqnarray}

Then, the distance between distant vertices can also be defined as

\begin{eqnarray}
  d(v_i,v_j; X^*) = \sum_{k  \in E_{i,j}} d(e_k, X^*) \label{eq:dd}
\end{eqnarray}

\noindent  where $E_{i,j}$ is a set of edges between vertices $V_i$ and $V_j$, and $X^*$ is the target vector. This distance depends on $X^*$, and varies according to the selected target vector. An example is discussed in Section \ref{sec:Application}, Applications. 

\subsection{Graph similarity}
Similarity between graphs can also be defined using cosine similarity as

\begin{eqnarray}
  S(G_m,G_n) = \frac{1}{N_{E_{m,n}}} \sum_{l  \in E_{m,n}} cos(X(e_{m,l}), X(e_{n,l})) \label{eq:gs}
\end{eqnarray}

\noindent  where $E_{m,n}$ is a set of corresponding edges in graphs $G_m$ and $G_n$, $N_{E_{m,n}}$ is the cardinality of the set $E_{m,n}$, and $e_{m,l}$ means the edge $l$ in the graph $G_m$. An example is discussed in Section \ref{sec:Application}, Applications. 

\section{Applications}
\label{sec:Application}
\subsection{Translation}

In this section we present some examples of the translation, edge distance and graph distance introduced above. A pre-trained word vector model named "word2vec-GoogleNews-vectors" which studies a 3-billion word Google News corpus is used in all the examples in this paper \cite{MM,LU,TS}.

\begin{figure*}
\begin{center}
\resizebox{\linewidth}{!}{\includegraphics{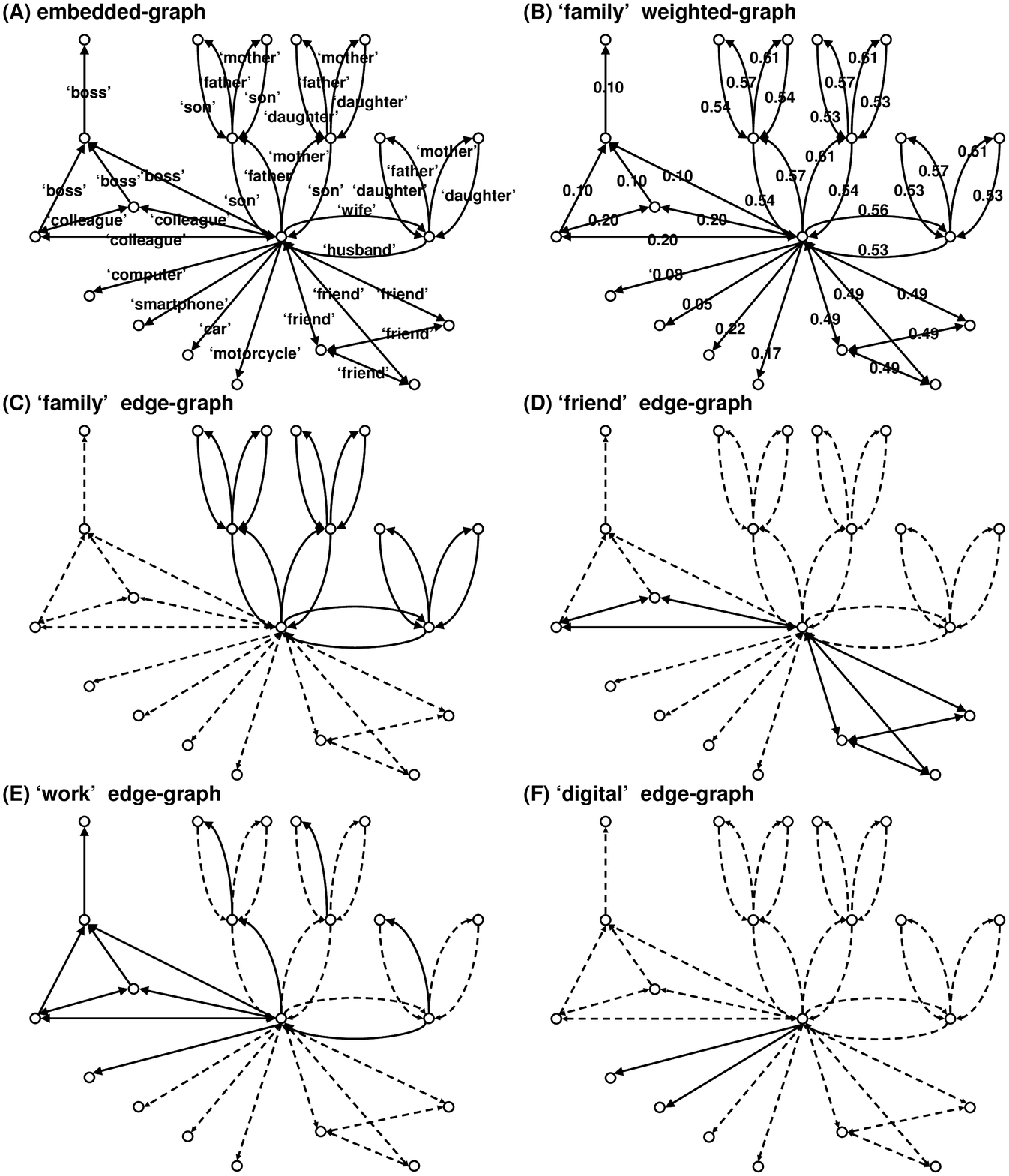}}
\end{center}
\caption{Examples of embedded-graph, weighted-graph and edges-graphs translated from the embedded-graph. (A) Original embedded-graph, (B) 'family' weighted-graph translated from (A), (C) 'family' edge-graph translated from (B), (D) 'friend' edge-graph translated from (A), (E) 'work' edge-graph translated from (A), (E) 'digital' edge-graph translated from (A).}
\label{F4}
\end{figure*}

An example of an embedded-graph is shown in Fig. \ref{F4}(A), in which the vertices stand for humans or objects, and the edges represent relations or roles. There is a main person at the center of the figure, who is linked to 'father', 'mother', 'boss', or other persons and objects; also, the father is linked to another 'father' who is the grandfather of the main person. All these relations or roles are expressed by vectors of word2vec, as shown in Table \ref{tb:T1}. 

First, we present an example of the translation. The embedded-graph (A) is translated into a weighted-graph by the vector 'family' and all edges are calculated using Eq. \ref{eq:cos} with $X^*_{'family'} = {0.11, 0.23, ..., 0.343}$. The corresponding weights of all the edges are shown in the column 'family' of Table \ref{tb:T2} and the 'family' weighted-graph created from the embedded-graph (A) is shown in Fig. \ref{F4}(B). This is an example of translation from an embedded-graph to a weighted-graph. Then, the weights of all edges are calculated by a threshold calculation with threshold set to $w_{'family'} = 0.5$. Specifically, if the weight of an edge is greater than $w_{'family'}$, then the edge is preserved, otherwise it is deleted. In this manner, the 'family' edge-graph is obtained from the 'family' weighted-graph, as shown in Fig. \ref{F4}(C), in which the preserved edges are depicted as plain lines, and the deleted ones are depicted as dotted lines. As shown in this figure, only the 'family' relations are preserved in the edge-graph, and others are deleted. The above procedure is used to translate from a weighted-graph to an edge-graph. The value of threshold  is determined manually. 

The obtained weights in Fig. \ref{F4}(B) are in similarity to the 'family' vector, because all edges of figure (A) are calculated using the cosine similarity to the 'family' vector given in Eq. \ref{eq:cos}. The weight becomes larger if the relation vector has a lot of 'family' vector components such as 'father' or 'mother', whereas it becomes smaller if the relation vector has few 'family' vector components. Then, edges with a lot of 'family' vector components are preserved and edges with few 'family' vector components are deleted by the threshold calculation with a proper threshold value.

This is the method to translate from an embedded-graph to a weighted-graph and an edge-graph. It is also the method to extract the edges related to any given target relation and to create the corresponding target relation edge-graph. Other instances are the following ones. The case where the target vector is 'friend' and the threshold $w_{'friend'} = 0.6$ is is shown in Fig. \ref{F4}(D), where the 'friend' edge-graph is created. The relation 'colleague' is also preserved as well as the relation 'friend'. This can also be seen in the real world as 'colleague' is a closer relationship than 'boss', and sometimes it becomes 'friend', that is, 'colleague' is relatively similar to 'friend'.

The case where the target vector is 'work' and the threshold $w_{'work'} = 0.1$ is shown in Fig. \ref{F4}(E). In this case, 'work' related edges such as 'colleague' or 'boss' are preserved, but 'father' or 'husband' and other non 'work' related edges are also preserved. In fact, we tried to extract only 'work' related relations such as 'colleague' and 'boss' , but the observed behavior occurs because the accuracy of the method totally depends on the accuracy of the used word2vec, so 'father' or 'husband' may be thought to be related to the word 'work', which is understandable, although it might be a stereotype. Furthermore, 'computer' is also preserved and this is also understandable because 'computer' is necessary for most 'work', recently.

The case where the target vector is 'digital' and the threshold $w_{digital} = 0.3$ is shown in Fig. \ref{F4}(F). Only digital products 'computer' and 'smartphone' are preserved.

\begin{table}
\caption{Word vectors used in Fig. \ref{F4} (A) embedded-graph}
\label{tb:T1}
  \begin{tabular}{| c || c | c | c | c |} \hline
    word & $x_1$ & $x_2$ & ... & $x_{300}$ \\ \hline
    mother & -0.0092 & -0.21 & ... &  -0.0077 \\ \hline
    father & 0.047 & -0.032 & ... &  0.088 \\ \hline
    ... & ... & ... & ... &  ... \\ \hline
    motorcycle & 0.14 & -0.027 & ... &  -0.25 \\ \hline
  \end{tabular}
\end{table}

\begin{table}
\caption{All the weights calculated from Fig. \ref{F4}(A) embedded-graph}
  \label{tb:T2}
  \begin{tabular}{| c || c | c | c | c |} \hline
    & family & friend & work & digital \\ \hline \hline
    mother & 0.61  & 0.54  & 0.09  & 0.02  \\ \hline
    father & 0.57  & 0.56  & 0.12  & 0.00 \\ \hline
    son & 0.54  & 0.54  & 0.04  & 0.02  \\ \hline
    daughter & 0.53  & 0.53  & 0.05  & 0.05  \\ \hline
    wife & 0.56  & 0.55  & 0.08  & -0.03  \\ \hline
    husband & 0.53  & 0.54  & 0.13  & 0.00  \\ \hline
    friend & 0.49  & 1.00  & 0.09  & -0.07  \\ \hline
    boss & 0.10  & 0.29  & 0.12  & 0.02  \\ \hline
    colleague & 0.20  & 0.62  & 0.11  & -0.04  \\ \hline
    computer & 0.08  & 0.08  & 0.19  & 0.37  \\ \hline
    smartphone & 0.05  & 0.04  & 0.02  & 0.35  \\ \hline
    car & 0.22  & 0.23  & 0.09  & 0.04  \\ \hline
    motorcycle & 0.17  & 0.18  & 0.04  & 0.02 \\ \hline
  \end{tabular}
\end{table}

\subsection{Edge distance}
\begin{figure}
\begin{center}
\resizebox{\linewidth}{!}{\includegraphics{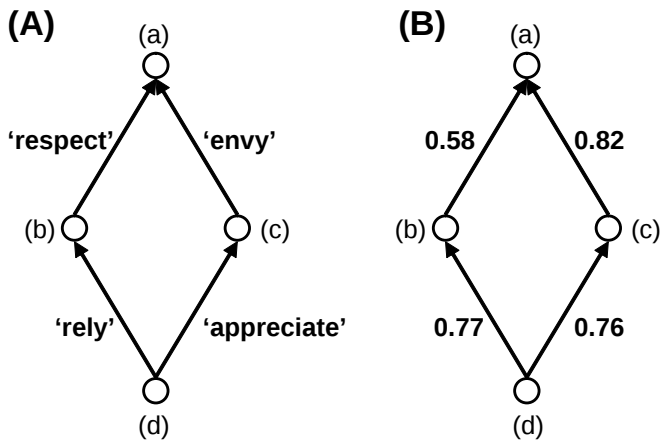}}
\end{center}
\caption{Example of distances between vertices of an embedded-graph. (A) original embedded-graph. (B) weighted-graph translated from (A) with the 'trust' relation.}
\label{F6}
\end{figure}

In this section we present an example of edge distance defined by Eq. \ref{eq:dd}, as shown in Fig. \ref{F6}. Fig. \ref{F6}(A) depicts the embedded-graph in which any vertex and edge stand for a person and an emotion toward a person, respectively person (d) relies on person (b) and appreciates person (c), person (b) respects person (a), and person (c) envies person (a). Now suppose that (a) has a business task to ask (d), but because (a) does not know (d) directly, then (a) has to ask (b) or (c) to ask (d) to do it. If the embedded-graph about the relations among the persons is considered, it would be possible to determine with word2vec whether person (a) should ask (b) or (c).

In this case, we choose the 'trust' vector to evaluate the route between (a) and (d), because 'trust' is an important trait in business relations. The words of all edges are translated into weights according to Eq. \ref{eq:d} using the 'trust' vector. Distances of each edge to 'trust' vector components are the following: 'rely' is 0.77, 'appreciate' is 0.76, 'respect' is 0.58, and 'envy' is 0.82. Accordingly, the route from (a) to (d) via (b) is 1.35, and the route from (a) to (d) via (c) is 1.58. Consequently, the distance of the route via (b) is shorter if we consider the point of view of 'trust', and therefore it is more appropriate, since the other route includes the emotion 'envy', which has smaller 'trust' vector components.

\subsection{Graph similarity}
\begin{figure*}
\begin{center}
\includegraphics[height=6cm,keepaspectratio]{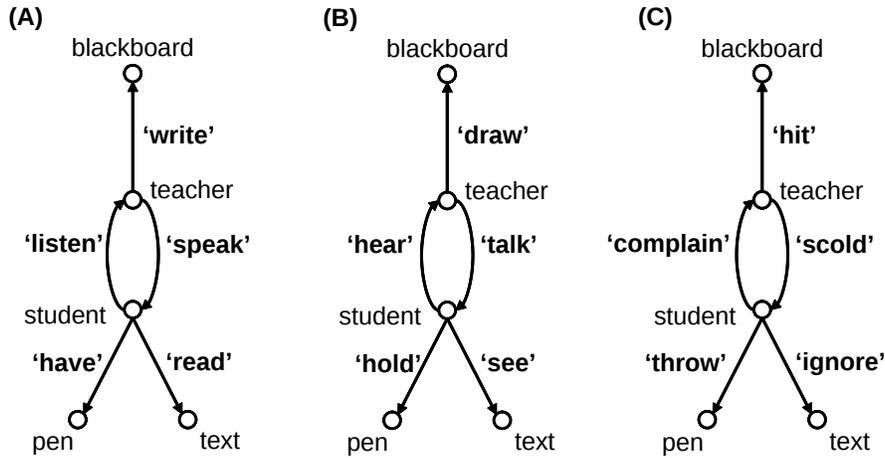}
\end{center}
\caption{Each figure (A), (B) and (C) shows a different embedded-graph. Plain texts stand for names of vertices and bold texts represent embedded vectors of edges.}
\label{FA}
\end{figure*}

In this section, we present an example of distance between embedded-graphs  as depicted in Fig. \ref{FA}. Each figure (A), (B) and (C) shows a different embedded-graph of a school class situation. A teacher speaks to students and writes something on a blackboard, students listen to the teacher, read a text and have a pen. Graph (A) and graph (B) are almost the same situation but graph (C) is very different and disrupted. The teacher scolds students and hits a blackboard, and students complain to the teacher, ignore a text, and throw a pen. The degree of difference or similarity can be calculated using Eq. \ref{eq:gs} in the embedded-graph.

The distances of corresponding edges in the two graphs are computed, for instance, the distance between 'speak' of (A) and 'talk' of (B) is 0.39. All distances of edges are computed and their average is the similarity between (A) and (B). The three distances between (A), (B) and (C) are shown in Table \ref{T3}. It can be seen that (A) and (B) are closer and (C) is farther from the others, which confirms our intuition.

In an edge-graph all edges represent the same relation, and a weighted-graph cannot deal with complicated relations such as those between words. Instead, an embedded-graph treats relationships and similarities between words such as 'speak', 'talk', or 'scold'. Accordingly, each edge of the graph has such a linguistic information, therefore the degree of similarity between relations can be obtained.

\begin{table}
  \caption{Graph similarity between the graphs depicted in Fig. \ref{FA}}
  \begin{tabular}{| c | c | c |} \hline
    (A)-(B) & (A)-(C) & (B)-(C) \\ \hline
    0.43 & 0.25 & 0.29 \\ \hline
  \end{tabular}
  \label{T3}
\end{table}


\section{Data structure}
\label{sec:Data}
There are three ways to represent an embedded-graph: adjacency matrix, edge list, and vector-labeled edge list.

\begin{figure*}
\begin{center}
\includegraphics[height=4cm,keepaspectratio]{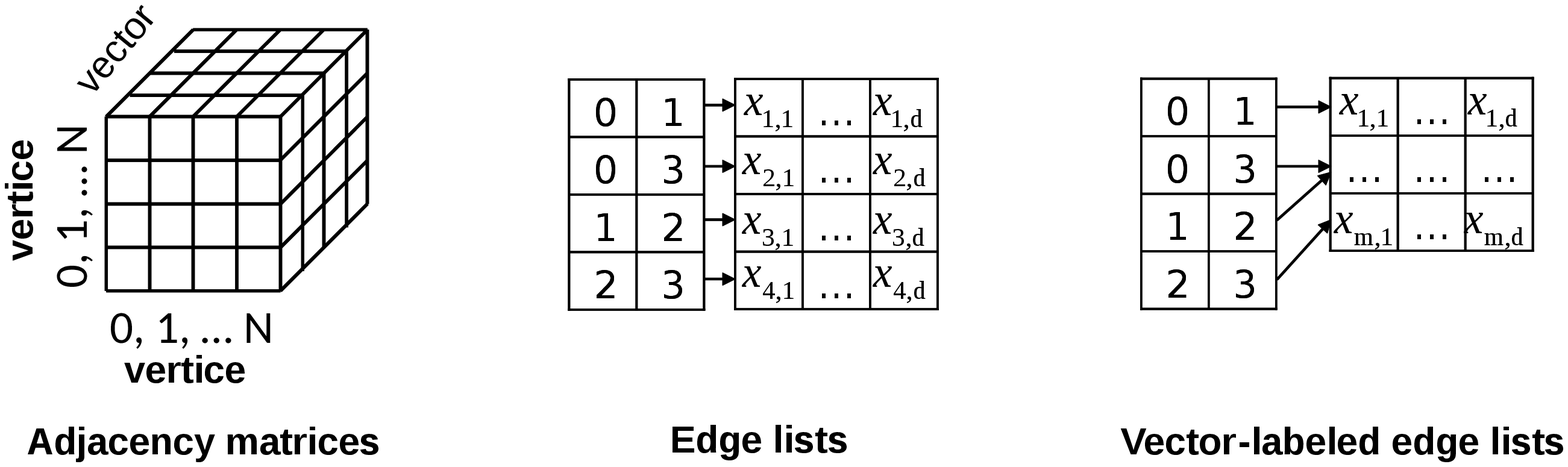}
\end{center}
\caption{Data structures for embedded-graphs: (A) adjacency matrix, (B) edge list, (C) vector-labeled edge list.}
\label{F3}
\end{figure*}

Adjacency matrix is shown in Fig. \ref{F3}(A), where the X and Y dimension stand for source and target vertices in the case of a directed graph. The Z dimension in the figure expresses the embedded vector, or the word vector in the case of word2vec. The presence of an embedded edge can be determined in constant time, but it takes $\mathcal{O}(N_v^2)$ space.

Edge list is shown in Fig. \ref{F3}(B). The first two columns on the left denote source and target vertices in the case of a directed graph. The right columns represent the embedded vector, or the word vector in the case of word2vec. The total space needed to store the graph is lower since it is equal to $\mathcal{O}(N_v)$.

Vector-labeled edge list is shown in Fig. \ref{F3}(C). Its structure is almost the same as the edge list except for the sharing vectors. The first two columns on the left stand for source and target vertices in the case of a directed graph, just as in an edge list. The third column is a link to the corresponding embedded-vectors. The total space is lower than the one needed for the edge list, but also the capacity of expression is lower, and it is not possible to modify vectors just for some specific edge.

\section{Discussion}
\label{sec:Discussion}
In this paper, we proposed embedded-graphs and their theory, which uses a distributed representation to describe the relations on the graph edges. Accordingly, embedded-graphs can express linguistic and complicated relations, which cannot be expressed by an edge-graph or a weighted-graph.

In Section \ref{sec:Theory},  we presented the mathematical definition of embedded-graph, translation, edge distance, and graph similarity. The translation method is the method which transforms an embedded-graph into a weighted-graph, and a weighted-graph into an edge-graph. By this method we have a theoretical consistency and the relations between the graphs become clear.

We also consider the distance between edges in an embedded-graph in a similar way as in the other types of graph. In this paper, we presented one instance of such distances, which is computed from the weighted-graph that is generated by the given embedded-graph using cosine similarity with a vector. This distance varies depending on the vector used for the cosine similarity computation.

The similarity between graphs is also defined and even if the network structure is completely the same, the similarity varies depending on the relations between vertices as shown in Fig. \ref{FA}. If their relations are similar, the similarity becomes greater, whereas if their relations are different, the similarity becomes smaller.

There would be other definitions of edge distance and graph similarity, but the definitions we introduced above are simple, and a strict theoretical discussion is beyond the purpose of this paper, which is just to introduce a new idea of graph.

There are labeled-graphs, where each edge is labeled with words, so they look just the same as embedded-graphs, but the labels are described as flags or scalar numbers in the computer data structure, and hence the linguistic meaning vanishes; for instance, 'speak', 'talk', and 'scold' are equally different words, although 'speak' should be closer to 'talk' in the linguistic meaning \cite{BU,KM}.

However, there are also some drawbacks in the concept of embedded-graphs as explained below.

\subsubsection{Development methods}
We presented some embedded-graphs in this paper such as in Fig. \ref{F1}(C), Fig. \ref{F4}(A), Fig. \ref{F6}(A), Fig. \ref{FA}, but those are all constructed manually. There is no algorithm or procedure to create an embedded-graph from data automatically. There might be three ways to realize the graph: (1) verbalize relations, then obtain vectors from word2vec or some other vector representations, (2) infer each vector component of the relations since each vector component of word2vec has a specific meaning, and therefore it can be inferred by regression method of machine-learning, (3) encode relations without word2vec, with an encoding algorithm, then obtain the relation vectors \cite{MW,ME,ME02}. The research on the development of embedded-graphs shall be treated in a future work.

\subsubsection{Accuracy}
Embedded-graphs use a distributed representation, which is often word2vec, and some computations are shown in Section \ref{sec:Theory}, Theory, based on cosine similarity, in the same way as most of the investigation related to word2vec \cite{MW,CH}. Accordingly, at the moment, the accuracy of an embedded-graph totally depends on the accuracy of word2vec. For example, cosine similarity between 'like' and 'trust' is 0.14, which is lower than 0.18, the similarity between 'hate' and 'trust'. We expected that a positive feeling such as 'like' would be based on 'trust', but actually, a negative feeling like 'hate' has more 'trust' components than 'like'. Accordingly, we need to improve the accuracy of word2vec in order to improve the accuracy of the embedded-graph.

Embedded-graphs can express many types of relations because they can represent all linguistic relations, so a lot of data can be encoded into one single graph, such as human-human relations, human-object relations, or object-object relations, and more. It means that embedded-graphs are better suited for big-data since they can represent a lot of data. Moreover, the information volume is also larger because most of the time word2vec has a lot of-column vectors, with a dimension such as 300 \cite{MM}. Machine learning or deep learning are useful to analyze embedded-graphs with such a massive load of data.

In this paper, we proposed a new type of graph, called embedded-graph, and we hope to improve it further both on the theoretical and the practical side.

\section{Conclusions}
In this paper, we proposed an embedded-graph and its theory, which employs a distributed representation to describe the relations on the graph edges. Embedded-graphs can express linguistic and complicated relations. We also introduced the mathematical definition of embedded-graph, translation, edge distance and graph similarity. Translation method is the method which transforms an embedded-graph into a weighted-graph. Edge distance of an embedded-graph is a distance based on the components of a target vector, and is computed using the cosine similarity with the target vector. Graph similarity is obtained considering linguistic complexity of the relations between words. We presented some examples and data structures in this paper as well.

\begin{acknowledgments}
We would like to show our gratitude to imatrix corporation for providing the opportunity to think about the first idea of embedded-graph.
\end{acknowledgments}

\end{document}